\documentclass[9pt,twocolumn,twoside]{ai_agents_journalism}
\usepackage{lmodern}
\usepackage[T2A,T1]{fontenc}
\usepackage[utf8]{inputenc}
\usepackage[russian,english]{babel}

\title{How can AI agents support journalists' work? An experiment with designing an LLM-driven intelligent reporting system}

\author[a]{Vasileios Maltezos}
\author[b]{Roman Kyrychenko}
\author[c]{Aleksi Knuutila}

\affil[a]{University of Helsinki}
\affil[b]{University of Helsinki}
\affil[c]{University of Helsinki}

\leadauthor{Maltezos}

\significancestatement{
This paper explores the deployment of agentic large language model (LLM) tools to support journalist reporting, with a focus on information discovery, summarization, and reporting workflows in conflict journalism. We integrate insights from journalist interviews and a case study of an LLM-powered reporting tool, examining both the technology's advantages and persistent challenges around verification, platform coverage, and ethical use in newsrooms.
}

\correspondingauthor{Corresponding author's email: vasileios.maltezos@helsinki.fi}
\begin{abstract}
The integration of artificial intelligence into journalistic practices represents a transformative shift in how news is gathered, analyzed, and disseminated. Large language models (LLMs), particularly those with agentic capabilities, offer unprecedented opportunities for enhancing journalistic workflows while simultaneously presenting complex challenges for newsroom integration. This research explores how agentic LLMs can support journalists' workflows, based on insights from journalist interviews and from the development of an LLM-based automation tool performing information filtering, summarization, and reporting. The paper details automated aggregation and summarization systems for journalists, presents a technical overview and evaluation of a user-centric LLM-driven reporting system (TeleFlash), and discusses both addressed and unmet journalist needs, with an outlook on future directions for AI-driven tools in journalism.
\end{abstract}

\dates{This manuscript was compiled on \today}
\doi{\url{http://ai-agents-journalism.org/}}

\begin{document}
\maketitle
\thispagestyle{firststyle}
\ifthenelse{\boolean{shortarticle}}{\ifthenelse{\boolean{singlecolumn}}{\abscontentformatted}{\abscontent}}{}

\dropcap{T}he integration of artificial intelligence into journalistic practices represents a transformative shift in how news is gathered, analyzed, and disseminated. Large language models (LLMs), particularly those with agentic capabilities, offer unprecedented opportunities for enhancing journalistic workflows while simultaneously presenting complex challenges for newsroom integration. An AI LLM Agent functions as a computational partner leveraging language models to interact in remarkably human-like ways. These digital assistants excel at understanding natural language and generating appropriate responses, enabling them to engage in conversations, deliver information, and support complex decision-making processes in journalistic contexts \citep{yang2024human}. The potential of these agents stretches well beyond basic text generation as they can function with increasing autonomy by completing assigned reporting tasks, gathering information from diverse sources, and synthesizing knowledge in ways that complement human journalists' capabilities \citep{ramos2025review}. This research explores the evolving landscape of automated journalism, with a particular focus on how these agentic LLMs can support journalists' workflows based on insights from interviews and the development of an LLM-based automation tool (TeleFlash) for journalists that performs information filtering, summarization, and reporting.

Automation in news discovery and delivery has established itself as a centrepiece of today's media information ecosystem, transforming how content gets filtered, aggregated, and condensed. Four key themes dominate this field: technological advancement, organizational change, ethical ramifications, and LLMs' growing significance. These elements reflect the evolving interplay between computational innovation and journalistic method, affecting everything from workflow efficiency to professional credibility. Technical advancements in natural language processing and generative AI form the foundation of automated journalism, with systems like natural language generation enabling rapid content creation for data-driven reporting, exemplified by Associated Press's use of automated earnings reports \citep{caswell2018automated}. Recent developments in LLMs, such as OpenAI's GPT models, expand capabilities beyond templated outputs, allowing nuanced text generation for tasks like news summarization and data analysis \citep{mishra2024use, smith2024ten}.

The rise of sophisticated language models marks a transformation in how journalists and machines can collaborate. These advanced systems offer unprecedented support for information collection, analysis, and presentation, yet their implementation presents substantial obstacles that must be addressed to maximize benefits while protecting journalistic principles. Ensuring factual accuracy and reliability remains paramount when developing LLM-powered news systems. For instance, investigations into the role of AI in journalism have highlighted concerns regarding algorithmic biases and the potential for these models to propagate inaccuracies if trained on uncurated data \citep{ademola2024ai, sonni2024bibliometric}. Thus, news organizations must develop robust fact-checking mechanisms alongside AI tools to ensure the trustworthiness of automated outputs.

Our research investigates the nuanced relationship between autonomous language models and journalistic practice, focusing on how these technologies enhance reporting processes while upholding ethical principles. We examine the development and implementation of TeleFlash, an LLM-driven reporting system that aggregates, filters, and summarizes Telegram posts from multiple channels on targeted topics. The system is designed specifically to address the challenge of discerning newsworthy insights on social media, enabling the creation of reports that cite sources of information and provide relevant statistics. Through analysis of journalist interviews and iterative adaptation of the program to their demands, we identify key considerations for the effective integration of AI tools in newsrooms, and which of them TeleFlash satisfies, and to what extent. The research is structured to provide a comprehensive literature review of automated journalism, followed by an in-depth case study of TeleFlash, an analysis of journalists' needs and satisfaction. Through this review and case study, we intend to understand, both in theory and practice, the needs of journalists and the possibilities of AI to cover these and other emerging needs in journalistic workflows. The main aim of the study is to explore the current possibilities of AI as a productive and assistive tool, as well as future directions for agentic LLMs in journalism.

\section*{Generative AI and Automation Tools: An Assistant, Not a Journalist}

The news industry increasingly incorporates automated systems into daily operations, particularly through advanced language models. This section examines current research on automation in journalism, focusing on technology implementation, newsroom adoption strategies, ethical considerations, and how agentic language models support journalistic workflows.

Automated journalism encompasses algorithmically generated news content using AI technologies, primarily relying on natural language generation. These systems enable customized content filtering, information aggregation, and report creation with minimal human oversight. Contemporary research highlights how language models are transforming journalism through unprecedented speed and content scalability improvements. The ability of models such as GPT-4 to generate coherent narratives from structured data is influencing how news organizations approach reporting \citep{hou2024benchmarking, wang2024thematic}. The automated generation of summarization reports based on data feeds, such as sports results and financial earnings, demonstrates the efficacy of LLMs in uncovering insights swiftly that journalists could use for the faster production of more comprehensive and informative articles.

Despite promising technological advances, newsrooms face significant implementation challenges. Research indicates that newsroom integration often encounters resistance from traditional journalists, concerns over job displacement, and the necessity for skills development related to AI technologies \citep{omar2024socio, oami2024performance}. For example, \citet{patel2024identification} articulates how the emotional dimension of journalistic narratives is challenging for LLMs to replicate, suggesting that while these models perform well on structured tasks, they may fail to convey the emotional depth characteristic of human journalism adequately. This limitation raises important questions about authenticity and human connection in machine-generated content.

Newsroom adoption of AI technologies remains uneven, shaped by organizational culture and perceived utility. While AI tools streamline workflows, enabling data journalists to identify trends or personalize content, integration challenges persist \citep{diakopoulos2019automating}. A survey of 292 journalists highlights shifting skill requirements, including prompt engineering and AI literacy, to leverage tools effectively \citep{reiter2024real}. Resistance often stems from fears of job displacement, though roles increasingly demand hybrid competencies in data interpretation and editorial oversight \citep{hair2024leveraging}.

Beyond technical considerations, ethical implications demand serious attention as these systems become more prevalent in journalism. The proliferation of generative AI in media brings concerns about misinformation, bias, and potential manipulation of public perception. Researchers have explored socio-demographic biases in AI applications, raising alarm bells when journalists rely on LLMs for nuanced topics requiring sensitivity \citep{omar2024socio, sekwatlakwatla2024evaluations}. Furthermore, the lack of transparency regarding the data sources and algorithms used to train these models exacerbates challenges in accountability and trustworthiness in news reporting \citep{patel2024identification}.

The integration of AI in journalism has brought to the forefront a range of critical ethical considerations that demand scrutiny. These challenges span issues of data bias and privacy, transparency and accountability, the impact on objectivity and independence, the potential for job displacement, and the overarching need for appropriate regulation and ethical guidelines. Using AI-generated content without proper citation constitutes plagiarism, aligning with established academic standards that prioritize originality and integrity. The risks of unethical use are significant, as highlighted by \cite{assad2024exploring}, who advocate for training focused on ethical AI usage to mitigate plagiarism concerns. Additional research indicates that AI-generated content may lack the critical thinking and creativity inherent in human-generated works, potentially undermining research reliability \citep{willie2024identifying, shopovski2024generative}.

Data bias stands as another significant ethical concern in the adoption of AI within newsrooms \citep{al2024artificial}. When AI systems learn from datasets containing societal prejudices, they risk perpetuating or amplifying these biases in their reporting, potentially resulting in unfair or discriminatory portrayals. Content personalization also raises serious privacy considerations regarding data collection and storage practices. Proactive safeguards must be established to minimize these potential harms and ensure responsible AI use in journalism.

The growing influence of automation in news production necessitates reconsidering fundamental concepts of authorship and journalistic principles. As automated tools produce content, the line distinguishing human-generated narratives from machine-generated articles blur, necessitating a reevaluation of what it means to be an author in this new landscape \citep{nam2024who}. Questions about attribution rights, plagiarism detection, and transparent disclosure of AI involvement must be addressed, particularly as research emphasizes the importance of responsible LLM usage guidelines.

Regarding practical applications, agentic language models have emerged as valuable workflow accelerators in newsrooms. For example, LLMs can assist journalists by evaluating articles, suggesting headlines, or summarizing lengthy reports, consequently streamlining the editorial process \citep{ross2024foundation}. Researchers argue that such integration not only improves efficiency but also allows journalists to focus on higher-level creative tasks, like investigating primary sources and focusing on the quality of their article texts, leading to more robust journalistic practices \citep{oami2024performance}. The arrival of autonomous AI systems presents multifaceted opportunities and challenges for contemporary media. Through integration of sophisticated AI technologies, communication processes are evolving in ways that transform journalistic methods, enhance audience engagement, and facilitate information sharing. Natural language processing and machine learning form the foundation of this shift, enabling AI systems to generate and analyze text with impressive skill. Reports emphasize that these conversational agents are adept at tailoring interactions based on user queries, thereby improving the relevance and personalization of communication \citep{hirome2024missional}. This technological capability is further substantiated by Liao et al., whose research highlights that AI can augment communication by enhancing positive sentiment in exchanges, particularly among experienced agents who leverage AI effectively \citep{liao2024effective}.

Human-machine collaboration in news reporting constitutes a rapidly expanding research and practice area. AI systems show tremendous potential for enhancing operational efficiency and content creation processes. Automated data processing capabilities facilitated by LLMs can mitigate the time constraints often faced by journalists, allowing them to focus more on investigative and creative storytelling \citep{gbaden2024challenges}. This integration can significantly improve news production timelines, ultimately contributing to a more responsive and informed public. Best practices for adopting LLM-driven reporting systems require a multifaceted approach that addresses technological challenges, mental health concerns, transparency, ethical governance, and educational initiatives. For instance, establishing interdisciplinary collaboration between journalists and AI experts can enhance the development and operationalization of LLMs in ways that prioritize both efficiency and ethical responsibility \citep{amponsah2024navigating}.

With this theoretical background in mind, we developed an LLM-based automation tool that is user-centric, implementing advancements tailored to journalists' needs, as expressed in a case study on monitoring and summarizing news about the Ukraine-Russia war.

\section*{TeleFlash: A Case Study in LLM-Driven Aggregation and Summarization Systems for Journalists}

Rather than beginning with code or implementation, our approach started by systematically mapping the practical requirements of journalists working under pressure, particularly those navigating the flood of social media content in conflict reporting. Through a focused set of interviews, we sought to surface concrete workflow challenges, habitual verification practices, and genuine frustrations, aiming to ground any later automation in lived professional experience rather than technological possibility alone. This needs analysis informed the entire development process, serving as the foundation for building TeleFlash, a reporting tool constructed to filter, summarize, and deliver actionable updates directly into a Slack workspace structured for clarity and feedback. Once the system was ready, the final preparatory step brought journalists into this environment, giving them access to automated summaries and, crucially, inviting their critical engagement with both the product and its fit within their established routines.

\subsection*{Interviewing journalists to understand their needs}

We conducted twelve semi-structured interviews with journalists from a range of Finnish and international media outlets, focusing mainly on those reporting about the Ukraine war. Participants included foreign correspondents, news desk editors, specialized reporters, and freelancers, providing a broad perspective on how social media shapes day-to-day journalism. Each interview followed a set structure but allowed for follow-up questions, targeting topics like social media use in newsgathering, verification problems, training needs, and the kinds of automation that might actually help in their work.

Telegram stood out as the main source for real-time updates, especially for stories developing in conflict zones. Several journalists explained that they rely on Telegram for initial information because posts appear faster there than through traditional wires. At the same time, journalists highlighted a set of major pain points: the sheer amount of information is unmanageable without filtering, the risk of errors is high because much of the content is unverified, and language barriers—particularly not knowing Ukrainian or Russian—cause inefficiency and dependence on external tools or colleagues. Many admitted their organizations lack standard procedures for vetting social media content, which forces them to rely on what they called "common sense" approaches that aren't consistent across newsrooms.

When asked directly what technology or automation would help, most journalists wanted tools that could automatically filter newsworthy posts by region or topic, provide accurate and fast translation, group or summarize key developments, and highlight which sources are likely to be credible. However, none wanted a black box: they insisted any system must include original post citations or links so journalists could check the context themselves before publication. With these criteria in mind, we set out to develop TeleFlash in an effort to address, to the extent possible, the overload and inefficiency journalists described in these interviews.

\subsection*{Technical Overview of the tool}

The development of TeleFlash represents a concrete implementation of LLM technology designed specifically to address the challenges journalists face when reporting on conflicts through social media channels. This case study examines the system's architecture, technical implementation, integration with journalistic practices, and multilingual capabilities to provide insights into the practical application of AI in news reporting workflows.

TeleFlash's system architecture is built around a modular framework that enables automated monitoring and analysis of Telegram channels' content, as a platform frequently used for information dissemination in conflict zones. At its core, TeleFlash employs a combination of regex-based filtering mechanisms and LLM-powered summarization to extract relevant information from large volumes of social media content. The system architecture consists of four primary components: a data collection module that interfaces with the Telegram API to retrieve messages from specified channels; a filtering module that applies customizable regex patterns to identify relevant content; a summarization module powered by an LLM that generates concise reports from filtered messages; and a distribution module that shares these summaries through Slack for team collaboration.


The technical implementation leveraged Python's asynchronous processing capabilities for handling substantial message volumes efficiently. TeleFlash operated on a scheduled basis, running daily at 6:00 AM to collect and process the previous 24 hours of content. The data retrieval process monitored over 170 different Telegram channels, storing messages in a PostgreSQL database for subsequent analysis. The pattern-matching system utilized specialized regex expressions in multiple languages, stored in configuration files that journalists could modify without programming knowledge. For Finland-related monitoring, TeleFlash employed six distinct regex patterns covering English, Russian, and Ukrainian variants of terms like 'Finland', 'Finnish', \textcyrillic{'Финляндия'}, and \textcyrillic{'фінський'}.

For summarization, TeleFlash connected to OpenAI's GPT models through their API, providing filtered messages as context and requesting concise summaries highlighting key information. The implementation included carefully crafted prompt engineering with specific instructions directing the model to emphasize factual reporting, focus on newsworthy developments (government actions, economic updates, security matters), avoid speculation, maintain a neutral tone, and identify potentially significant information. Each summary included explicit message ID citations, enabling journalists to verify original sources. The system also translated summaries into Finnish, making the content accessible to non-Russian-speaking audiences.

TeleFlash's information-gathering design reflected an understanding of workflows, derived from the interviews, during conflict reporting. By automating routine monitoring of 170+ channels, it directly addressed the information overload problems frequently mentioned by conflict zone reporters. The system generated two comprehensive reports: one in English and one in Finnish, each containing the AI-generated summary with embedded source links and detailed engagement metrics. These metrics included view counts, forwards, channel distribution analysis, and virality scores—providing quantitative indicators of message reach and potential significance.

The system's implementation of source citation represented a particularly valuable feature that facilitated journalistic verification, even though it did not automate it. For each summarized point, TeleFlash included the original message ID as a citation, while the distribution interface transformed these IDs into clickable links to the original Telegram posts. This functionality directly addressed transparency requirements, allowing journalists to quickly verify claims against original sources and evaluate message credibility without manual searching.

Despite its capabilities, TeleFlash is currently focused exclusively on Telegram due to the case study and the fact that it was in a prototype testing phase. Its modular architecture could potentially expand to incorporate other platforms, enhancing its value for journalists monitoring multiple information channels simultaneously.

The system depended on human journalists for final verification, contextual understanding, and editorial judgment, reinforcing that LLM-driven tools are functioning most effectively when designed to enhance rather than replace journalistic expertise. Its case demonstrated both possibilities and constraints of current LLM-driven reporting systems.

\subsection*{Pilot Phase: Getting the journalists to test and evaluate TeleFlash}

From the outset, the pilot phase exposed steady hurdles in journalist participation and tool use. Out of more than thirty journalists invited by email, only eleven accepted the invitation and joined the dedicated Slack workspace. The onboarding process emphasized clarity: we prepared step-by-step written instructions, including a channel solely for general guidance and another for feedback. Despite these efforts, making the instructions truly intuitive proved difficult. Some journalists mentioned they were unsure how to begin, with a few admitting they skipped the instructions entirely or became stuck at basic steps.

A common barrier was the choice of Slack as the delivery platform. While Slack is popular in many industries, several journalists were either not regular users or were unfamiliar with how to navigate its channels and notifications. This unfamiliarity caused a noticeable delay: some journalists took several days or even weeks to join the workspace after receiving the email, while a handful joined but did not interact further. Those who did access the summaries often limited their activity to reading daily posts.

Our attempts to gather feedback met with limited response. Both in-Slack feedback channels (with explicit instructions) and a separate Google Forms survey failed to generate commentary, even though we could see the journalists were visiting the workspace and viewing the reports. The lack of input suggests that a tool must meet an immediate, clearly felt need to motivate active use and integration into established workflows. As we transitioned to technical evaluation, these pilot results highlighted the practical barriers that even well-designed newsroom automation tools must overcome to move from passive use to active, valuable adoption.

\section*{Insights from TeleFlash Implementation: Needs Addressed and Room for Improvements}

The development of TeleFlash as an LLM-driven reporting system was guided by insights gathered from extensive journalist interviews, focusing specifically on challenges encountered during conflict reporting via social media. This chapter presents a comprehensive analysis of how effectively TeleFlash addresses the needs expressed by journalists in their daily workflows.

Our evaluation indicated that TeleFlash successfully addressed several crucial requirements identified by journalists. The regex-based filtering system effectively extracts relevant content from large social media volumes, addressing information overload challenges. As one journalist noted during interviews, "The sheer volume of posts makes it impossible to track relevant developments manually" (Interview 2). TeleFlash's automated retrieval of messages from specified Telegram channels within a 24-hour timeframe provides the real-time monitoring capability that journalists identified as essential, particularly when covering rapidly evolving situations in conflict zones where traditional reporting access is limited.

The multilingual regex patterns implemented in TeleFlash (supporting English, Russian, and Ukrainian) partially address language barriers that journalists identified as significant obstacles when analyzing foreign sources. This functionality allows journalists to capture relevant content across languages, though the system lacks more sophisticated translation capabilities for specialized terminology. The AI-powered summarization feature generates concise reports tailored to filtered messages, addressing journalists' expressed need for tools that can efficiently distill key points from extensive datasets. As one journalist emphasized, "We need something that can give us the essence without having to read through hundreds of posts" (Interview 5).

TeleFlash's integration with Slack for sharing filtered summaries directly to team channels facilitates the collaboration that journalists identified as crucial for effective newsroom workflows. The inclusion of engagement metrics (views, forwards) in these reports provides additional context that helps journalists assess the potential impact and reach of specific content.

However, we also identified several areas where TeleFlash falls short of meeting the needs expressed by the journalists. While the system filters messages based on keywords and sources, it does not include robust verification mechanisms to assess source credibility or cross-check information, a capability repeatedly emphasized by journalists as essential for responsible reporting. The difficulty of choosing the right set of information sources remains a challenge, despite the plethora of AI and automation tools that process and filter information. As one journalist stated, "Distinguishing reliable sources from misinformation remains our biggest challenge" (Interview 4).

Ideally, a system should be smart about what is worth reporting and what isn't. We attempted to integrate this functionality into TeleFlash through prompt engineering, i.e., by instructing the system to evaluate whether the initially distinguished posts are worthy of being reported as "news". This approach, though, has a significant disadvantage: its efficiency can vary, and the logic based on which the decision of noting a post as "newsworthy" is made cannot be monitored. This can lead to various errors, such as false exclusions of newsworthy information or false inclusions of non-newsworthy information.

Factual mistakes can also happen due to a lack of context or hallucinations of the LLM. During the pilot phase, a mistake of this kind was noticed when the same person was mentioned in two posts that TeleFlash reported, but the report implied that it was a different person in each case.

From another perspective, TeleFlash's exclusive focus on Telegram channels without integration of data from other platforms like X (formerly Twitter) or Facebook limits its utility for journalists who need to monitor multiple platforms simultaneously. The system also lacks user-friendly guides or training modules that would help journalists unfamiliar with such tools to utilize them effectively, despite interview data suggesting that training in social media monitoring techniques is often insufficient.

\begin{table*}[!ht]
\caption{Summary of key journalist requirements and TeleFlash features}
\label{tab:journalist-needs}
\begin{tabular}{p{0.36\textwidth}p{0.10\textwidth}p{0.49\textwidth}}
\toprule
Journalist Request or Challenge mentioned & Addressed by TeleFlash? & Explanation \\
\midrule
Filtering relevant information & Yes & Regex-based keyword filtering extracts actionable content efficiently. \\
Monitoring real-time updates & Partially & Automates data retrieval from Telegram channels within the last 24 hours. \\
Multilingual content handling & Partially & Supports multilingual filtering but lacks robust translation capabilities. \\
Summarizing large volumes of data & Yes & AI-powered summarization provides concise reports tailored to topics. \\
Collaboration and sharing & Yes & Posts summaries directly to Slack with engagement metrics included. \\
Ethical use and verification of sources & Partially & Does not assess source credibility but focuses on public channels only. \\
Comprehensive platform integration & No & Focuses only on Telegram; no integration with other platforms like X. \\
Advanced language support & Partially & Includes advanced translation tools but lacks standardized terminology glossaries. \\
Training and usability improvements & No & Does not offer training resources for new users unfamiliar with tools. \\
Addressing data overload & Yes & Has features to rank or prioritize content based on newsworthiness. \\
Distinguishing sources of information based on their credibility & No & TeleFlash has no integrated mechanism to assess credibility. \\
\bottomrule
\end{tabular}
\end{table*}

\section*{Future Directions for Agentic LLMs in Journalism}

Examining TeleFlash's implementation and journalist response reveals important development paths for future agentic language model systems in journalism. These insights suggest more sophisticated, integrated approaches to supporting journalistic workflows through advanced AI.

A key improvement area involves developing more robust verification mechanisms along with mechanisms for selecting the information sources and distinguishing important ones. Future iterations should implement automated cross-checking across multiple sources, leveraging language models' reasoning capabilities to evaluate source credibility and identify potential misinformation. This would address a critical gap identified in our analysis, which is the journalists' need for tools that not only gather content but also help verify its reliability.

Perhaps most significantly, future systems should explore enabling autonomous language models to identify additional relevant sources based on content analysis. Rather than relying solely on predefined source lists, these systems could proactively discover potentially valuable information sources by analyzing content patterns and network relationships. This capability would help journalists uncover perspectives they might otherwise miss, particularly in rapidly evolving situations where new sources emerge quickly.

Multi-platform integration represents another essential development direction. Journalists consistently expressed frustration with monitoring separate platforms independently. Future systems should create unified interfaces aggregating content from Telegram, X, Facebook, and other relevant platforms, providing comprehensive coverage without requiring journalists to switch between multiple tools. Such integration would substantially improve newsroom efficiency.

Enhanced language support emerges as another critical improvement area. While TeleFlash implemented basic multilingual filtering, future systems should incorporate sophisticated translation capabilities, particularly for specialized terminology common in conflict reporting. This might include domain-specific glossaries and context-aware translation models that understand nuanced or technical language.

Refining scheduling algorithms offers promising avenues for enhancing autonomous language model capabilities in journalism. Future systems could intelligently determine optimal monitoring frequencies based on topic volatility or regional instability, automatically adjusting data collection parameters during breaking news events. This adaptive approach would ensure timely updates during critical periods while reducing information overload during quieter times. The ideal case would be to report immediately when something interesting happens. Nonetheless, this type of automatic monitoring and reporting is a challenging procedure to operationalize, not only due to technical difficulties but also due to the inherently subjective criteria of what is interesting, what is urgent to report, etc.

Lastly, feedback mechanisms are also an area where improvement is needed. Agents and LLMs may be inaccurate or fail to meet journalists' needs when conducting specific tasks; however, journalists often have distinct interests. How could they redefine and readjust the focus and the properties of the summaries or the information collection process in real time, according to their specific interests? This challenge can be addressed by implementing systems that allow journalists to rate the relevance and accuracy of AI-generated summaries, enabling continuous learning and model adaptation. The implementation of a robust mechanism for adequately utilizing feedback and making optimal adjustments is also, though, technically challenging. Nevertheless, such a feedback loop, if appropriately implemented, would progressively improve output quality while giving journalists greater control over how the system evolves to meet their specific requirements.

The development of these enhanced capabilities must be guided by ongoing dialogue with journalists to ensure that technological innovations align with practical needs. As one journalist noted in our interviews, "The tools need to work the way we work, not force us to adapt to them" (Interview 8). This human-centered approach to development will be essential for creating systems that genuinely enhance journalistic capabilities rather than adding technological complexity without corresponding benefits.

\section*{Conclusions and Thoughts about the Future}

Our examination of autonomous language models in journalism reveals a nuanced landscape of possibilities and challenges shaping tomorrow's news production. TeleFlash demonstrates that AI-driven tools can effectively address specific journalistic needs, particularly in high-pressure contexts like conflict reporting. However, our analysis also highlights current technological limitations and the enduring importance of human judgment in news production.

AI integration in journalism represents not replacement but reconfiguration of journalistic work. As demonstrated through TeleFlash, autonomous language models effectively automate routine monitoring, filter relevant information from overwhelming content volumes, and generate preliminary summaries serving as foundations for deeper analysis. This automation redirects journalists' attention toward activities demanding human judgment, contextual understanding, and ethical decision-making.

Journalist feedback analysis reveals that successful AI integration depends on tools designed with specific journalistic requirements in mind rather than imposing technological capabilities disconnected from practical workflows. TeleFlash's well-received filtering and summarization features, alongside criticisms of its limited verification capabilities and platform coverage, highlight the importance of ongoing dialogue between technology developers and working journalists. This collaborative approach ensures AI tools enhance rather than hinder journalistic practice.

Ethical considerations around AI in journalism require continuous attention as these technologies evolve. Issues of transparency, accountability, bias mitigation, and source verification remain inadequately addressed in current implementations. As autonomous capabilities advance, questions about algorithmic influence on news judgment gain importance. News organizations must establish clear policies for AI tool usage, ensuring technological innovation doesn't compromise journalistic integrity or audience trust.

Educational implications represent another critical consideration. Today's changing technological landscape requires journalists to develop new skills in areas like prompt engineering, output assessment, and algorithmic literacy. Journalism education must evolve to incorporate these competencies while reinforcing foundational principles of verification, accuracy, and ethical reporting that remain essential regardless of technological context.

Future development of autonomous language models in journalism will likely trend toward integrated systems combining multi-platform monitoring, enhanced verification capabilities, and sophisticated language processing. As these systems evolve, maintaining complementary relationships between human and machine capabilities remains essential for preserving journalism's social value in democratic societies. The journey toward effective human-machine collaboration in journalism has only begun. The challenges identified, from verification mechanisms to cross-platform integration, represent opportunities for innovation rather than insurmountable obstacles. By focusing on journalistic needs and values while leveraging technological capabilities, news organizations can navigate this transition in ways that strengthen journalism's essential role in public discourse.

\acknow{This research was funded by Media-alan tutkimussäätiö (Media Industry Research Foundation of Finland).}
\showacknow

\bibliography{references}

\section*{Appendix}

\section*{Prompts for Report Creation}

\subsection*{System Prompt}
\begin{quote}
You are an expert political analyst and journalist specializing in Finnish affairs capable of summarizing and finding commonalities in Russian-language messages about Finland or Finnish topics. Focus exclusively on newsworthy developments:

Major policy decisions and governmental actions

Economic and trade developments

Security and defense matters

Diplomatic relations

Infrastructure and strategic developments

Any other significant national developments

Exclude:

Cultural events

Social media discussions

Entertainment news

Human interest stories

Anecdotal mentions

Humor or entertainment

Writing requirements:

1. Write in clear journalistic style  
2. Always cite message IDs in parentheses within sentences  
3. Focus on factual reporting  
4. If only no newsworthy content exists, state (without making any summary): "Nothing newsworthy was mentioned the last day"  
5. Maintain neutral, objective tone
\end{quote}

\subsection*{User Prompt}
\begin{quote}
Analyze these Finland-related messages for newsworthy developments and produce a summary based on them:

{messages\_text}

If newsworthy content exists, structure your response as follows:

\textbf{Overview:}\\
Brief summary of key developments.

\textbf{Key Topics:}\\
Detailed coverage of significant developments, with each development in its own paragraph.

If no newsworthy content exists at all, simply state (without making any summary):\\
"Nothing newsworthy was mentioned the last day"

Requirements:

1. Focus on verified developments\\
2. Always cite message IDs in parentheses\\
3. Maintain professional writing style and neutral tone while attributing claims to sources\\
4. Group related developments together\\
5. Only include significant developments\\
6. Present information as channel claims using phrases like but not limited to:\\
\ \ "(message ID) reported that..."\\
\ \ "According to (message ID)..."\\
\ \ "Several channels (message IDs) claimed that..."
\end{quote}

\subsection*{Finnish Translation System Prompt}
\begin{quote}
You are a professional translator specializing in English to Finnish translation. Translate the text while preserving all formatting, numbers, and special characters. Keep message IDs in their original form.
\end{quote}

\begin{quote}
Translate the following text to Finnish:

{summary}
\end{quote}

\section*{Report Template Structure}
\begin{quote}
Finland-Related Messages Summary

Analysis Period: {date range} \\
Generated: {current time}

Analysis of Messages about Finland and Generated Summary:\\
------------------------------------------------------------\\
{Generated Summary}\\
------------------------------------------------------------

Basic Metrics:\\
- Total Messages Analyzed (related to Finland): {metrics['total\_messages']}\\
- Total Views: {metrics['total\_views']}\\
- Total Forwards: {metrics['total\_forwards']}\\

Average Metrics:\\
- Avg Views/Post: {metrics['avg\_views']} (average views per post)\\
- Avg Forwards/Post: {metrics['avg\_forwards']} (average forwards per post)\\
- Base Engagement: {metrics['engagement\_rate']}\% (share rate)\\

Advanced Engagement:\\
- Views/Forwards Ratio: {metrics['views\_to\_forwards\_ratio\_avg']}\\
- Virality Score: {metrics['virality\_score']}\%\\
- Unique Channels: {metrics['unique\_channels']}\\

Distribution Patterns:\\
- Posts/Day: {metrics['posts\_per\_day']}\\
- Peak Daily Posts: {metrics['max\_daily\_posts']}\\
- Channel Activity Ratio: ${metrics['total\_messages'] / metrics['unique\_channels']}$\\

Message IDs are clickable links to original posts.
\end{quote}

\end{document}